\documentclass[superscriptaddress,twocolumn]{revtex4}
\usepackage{amsmath}
\usepackage{amsfonts}
\usepackage{amssymb}
\usepackage{graphicx}

\begin{document}

\title{Local extraction of EPR entanglement from classical systems}
\author{D. Kaszlikowski}\affiliation{Departament of Physics, National University of Singapore, 117542 Singapore, Singapore}
\author{V. Vedral}\affiliation{The School of Physics and Astronomy, University of Leeds, Leeds LS2 9JT, UK}\affiliation{Departament of Physics, National University of Singapore, 117542 Singapore, Singapore}
\begin{abstract}
Coherent states with large amplitudes are traditionally thought of as the best quantum mechanical approximation of classical behavior.
Here we argue that, far from being classical, coherent state are in fact highly entangled. We demonstrate this by showing that a general system of indistinguishable bosons in a coherent state can be used to entangle, by local interactions, two spatially separated and distinguishable non-interacting quantum systems. Entanglement can also be extracted in the same way from number states or any other nontrivial superpositions of them. 
\end{abstract}
\maketitle

It has been experimentally demonstrated that Bose-Einstein condensates (BECs) behave in many respects like coherent light. Three examples of such a behaviour are an interference of two condensates \cite{BEC-interference}, diffraction of condensates from a beam of coherent laser light \cite{BEC-diffraction} and four-wave mixing of condensates \cite{bec-mixing}. To explain and model these phenomena we can simply assume that BECs are described by coherent states  of atoms \cite{glauber} pertaining to the ground state of an atomic trap \cite{coherent-BEC} from which they are released. The development of techniques allowing to extract a beam of coherent atoms from a BEC without disturbing it is very often referred to as an "atom laser" \cite{atom-laser} emphasising the analogy with coherent light. 

Properties of coherent states of light are very similar to that of classical monochromatic light with a definite phase and amplitude \cite{glauber}. Coherent states are, in fact, the best quantum approximation to classical coherent light. Therefore, if a coherent beam of atoms from a BEC can be described by a coherent state of matter waves, a natural question is to ask whether such a BEC contains any entanglement at all. The first impulse is to say that BEC is not entangled because, when the number of atoms in a trap is known, say $n$, and they do not interact, then their total state $|\Psi\rangle$ is a product of single particle ground states of the trap $|\phi\rangle$, i.e., $|\Psi\rangle = |\phi\rangle^{\otimes^n}.$ This clearly looks like a separable state, especially if we write it in the form of a wave function in the position representation $\Psi(\vec{r}_1,\vec{r}_2,\dots,\vec{r}_n)=\phi(\vec{r}_1)\phi(\vec{r}_2)\dots\phi(\vec{r}_n)$. This kind of a wave function, for large $n$, is often called a "super atom" to stress the fact that many seemingly independent atoms behave coherently like a single quantum object.

In this paper we will show that the above statement is incorrect and that the coherent states modeling the BEC as well as states of the type $|\Psi\rangle = |\phi\rangle^{\otimes^n}$ are spatially entangled. Our proof is based on the fact that this entanglement can be locally extracted in the form of an EPR-like pair of distinguishable quantum particles. We show that the extraction scheme produces a maximally entangled EPR pair in the classical limit (we will explain carefully what the classical limit is below) leaving the BEC almost undisturbed. Finally, we discuss a possible experimental implementation of the scheme.

We would like to note that there has been previous work discussing entanglement in BECs, see for instance \cite{previous}. What makes our approach different is that we present an operational method of extracting useful entanglement (quantum-information-processing-ready) without disturbing the BEC. The situation is somewhat similar to creation of a pair of entangled photons in the process of parametric down conversion where a crystal and a stimulating laser light are left almost intact after the whole process.

We now present our extraction scheme in the simplest possible way but without losing any essential features. We assume that we have a BEC consisting of spinless non-interacting bosons trapped in a one dimensional harmonic trap. The assumption about the dimensionality of the trap is also intended only to simplify calculations and our results can be easily generalized to two and three dimensional traps. Moreover, it has been shown in the Ref. \cite{ketterle} that BEC phenomenon can be also observed in one dimension and therefore even our simplified model is physically relevant.

Let us consider a spinless boson with mass $m$ in a harmonic oscillator potential with frequency $\omega$ coupled to two quantum probes called $L$ and $R$ via the following interaction Hamiltonian 
\begin{equation}
H_I(t) = g(t)\left(\theta_L(X)P_L+\theta_R(X)P_R\right),
\label{interaction}
\end{equation}
where $X$ is the position operator for the particle, $\theta_{L(R)}(x)$ is the characteristic function of the region $L=<-\infty,0> (R=<0,\infty>)$. $P_L$ and $P_R$ are momentum operators for quantum probes. The coupling strength $g(t)$ depends on time, which allows us to switch on and off the interaction between the particle and the probes. We assume that probes are spatially separated, distinguishable and non-interacting physical systems. Although the probes can in principle be realized by any physical system in this paper we assume that they are particles of mass $M$ in a harmonic oscillator potential with a frequency $\Omega$. 

The Hamiltonian (\ref{interaction}) describes a scenario when each quantum probe $L$ and $R$ interacts with the particle only if the particle is in the region $L$ or $R$ respectively. This is important since we want to demonstrate that entanglement present in the trap can be transferred locally to the initially disentangled and spatially separated probes.  

We are interested in the situation where there are many non-interacting indistinguishable particles (bosons, both massive and massless) and each probe consists of 
only one particle. Therefore it is convenient to write the operators $\theta_L(X)$ and $\theta_R(X)$ appearing in the formula (\ref{interaction}) in the second quantized form. There is no need to do this with the operators $P_L$ and $P_R$ as they act on single particles only. We have
\begin{equation}
H_I(t) = g(t)(\Lambda_L P_{L}+\Lambda_R P_{R}), 
\label{sec-int}
\end{equation}
where
\begin{equation}
\Lambda_{L(R)}=\sum_{k,l}\lambda^{(L(R))}_{k,l}a^{\dagger}_ka_l.
\end{equation}
Here $a_k$ denotes an anihilation operator of the orbital $\phi_k(x)$ in the harmonic oscillator potential of particles and 
$\lambda^{(L(R))}_{k,l}=\int dx\phi^{*}_k(x)\theta_{L(R)}(x)\phi_l(x)$, which is the overlap of orbitals $\phi_k(x)$ and $\phi_l(x)$ in the region $L(R)$. To simplify notation we assume that the orbitals are real functions. Note that $[\Lambda_L,\Lambda_R]=0,$
where we have used the formula $\sum_{m}\lambda^{(L)}_{km}\lambda^{(R)}_{mn}=\int dx\phi_k(x)\phi_l(x)\theta_L(x)\theta_R(x)=0$. The vanishing of the commutator is consistent with the requirement of local interaction of the particles with the probes.

Let us assume that at time $t=0$ the state of the particles and the probes is
\begin{equation}
|\psi(0)\rangle = |\phi\rangle|0\rangle_L|0\rangle_R,
\end{equation}
where $|0\rangle_{L(R)}$ is the ground state of the probe $L(R)$ and $|\phi\rangle$ is some arbitrary state of the particles. In the first order perturbation (see discussion below for the validity of this approximation) the state of the system after a time $T$ reads
\begin{eqnarray}
&&|\psi(T)\rangle = |\psi(0)\rangle-i[T\hat{H_0}|\phi\rangle|0\rangle_L|0\rangle_R-\nonumber\\
&&i g(\Lambda_L|\phi\rangle P_L|0\rangle_L|0\rangle_R+
\Lambda_R|\phi\rangle |0\rangle_L P_R|0\rangle_R)],
\end{eqnarray} 
where $T$ is the time of interaction and $g=\int_{0}^{T}dt g(t)$ (the above ket is not normalized). The operator $H_0$ describes the free evolution of the particles and probes. It does not have any effect on the initial states of the probes as they are stationary states but in general it affects the state $|\phi\rangle$. The effect of the action of momentum operator on the ground state of the harmonic oscillator of the probe $L(R)$ is $P_{L(R)}|0\rangle_{L(R)} = i\sqrt{\frac{M\Omega}{2}}|1\rangle_{L(R)}$, where $|1\rangle_{L(R)}$ is the first excited state of the probe $L(R)$. Therefore, we have
\begin{eqnarray}
&&|\psi(T)\rangle =(|\phi\rangle-iT\hat{H_0}|\phi\rangle)|0\rangle_L|0\rangle_R+\nonumber\\
&&g\sqrt{\frac{M\Omega}{2}} (\Lambda_L|\phi\rangle|1\rangle_L|0\rangle_R+
\Lambda_R|\phi\rangle |0\rangle_L|1\rangle_R).
\end{eqnarray}

If the scalar product $\langle\phi|\Lambda_R^{\dagger}\Lambda_L|\phi\rangle$ is non-zero the above state contains entanglement that can be recovered by performing a projective non-demolishion measurement on the probes checking if both of them are in the ground state or not. If they are we get a separable state of the probes $|0\rangle_L|0\rangle_R$ otherwise we get an entangled state (after tracing out the particles) 
\begin{eqnarray}
&&\rho = \frac{1}{S}\left(\langle\phi|\Lambda_L^{\dagger}\Lambda_L|\phi\rangle |1\rangle_L|0\rangle_R{}_L\langle 1|{}_R\langle 0|+\right.\nonumber\\
&&\left. \langle\phi|\Lambda_R^{\dagger}\Lambda_R|\phi\rangle |0\rangle_L|1\rangle_R{}_L\langle 0|{}_R\langle 1|+\right.\nonumber\\
&&\left. \langle\phi|\Lambda_L^{\dagger}\Lambda_R|\phi\rangle |1\rangle_L|0\rangle_R{}_L\langle 0|{}_R\langle 1|+\right.\nonumber\\
&&\left. \langle\phi|\Lambda_R^{\dagger}\Lambda_L|\phi\rangle |0\rangle_L|1\rangle_R{}_L\langle 1|{}_R\langle 0|\right),
\end{eqnarray}
with $S=\langle\phi|\Lambda_L^{\dagger}\Lambda_L|\phi\rangle+\langle\phi|\Lambda_R^{\dagger}\Lambda_R|\phi\rangle$. The negativity \cite{werner} $\mu$ of this operator is given by the simple formula $$\mu=\frac{|\langle\phi|\Lambda_L^{\dagger}\Lambda_R|\phi\rangle|}{S}.$$

Nothing much can further be said about $\mu$ unless we specify the state $|\phi\rangle$. We are interested in the entanglement in a coherent state of the bosons in the trap so we put $|\alpha\rangle = \exp{(-\frac{1}{2}|\alpha|^2)}\sum_{n=0}^{\infty}\frac{\alpha^n}{\sqrt{n!}}|n\rangle$, where $|n\rangle = \frac{1}{\sqrt{n!}}(a_0^{\dagger})^n|vac\rangle$. For this choice of $|\phi\rangle$
the negativity $\mu$ reads
\begin{equation}
\mu = \frac{1}{2}\left(\frac{|\alpha|^2}{2+|\alpha|^2}\right) = \frac{1}{2}\left(\frac{\langle N\rangle}{2+\langle N\rangle}\right), 
\end{equation}
where $\langle N\rangle$ is the average number of atoms in the trap. The negativity goes to its maximal value $\frac{1}{2}$ for $|\alpha|^2$ going to infinity. This is very surprising because the bigger the $|\alpha|$ the more classical the state $|\alpha\rangle$ is. We can explain this counterintuitive feature in the following simple way. To do so we compute the fidelity, $F$, between the initial state of BEC $|\alpha|$ and the disturbed final state $\Lambda_{L(R)} |\alpha\rangle$ (which has to be normalised):
\begin{equation}
F = \frac{1}{\sqrt{1+\frac{2}{|\alpha|^2}}}
\end{equation} 
It is clear that the disturbance decreases with the increase in amplitude of the coherent state. But this is exactly what leads to a higher amount of transfered entanglement to the probes since the BEC state factors out in the final state in the classical limit of $|\alpha| \rightarrow \infty$. Therefore, quantum states in the classical limit play a very intriguing dual role here; they generate entanglement by local means in the probes, but preserve their classicality by not becoming entangled with the probes.   

Equally interesting is the fact that the initial state of the BEC does not need to be coherent. 
If the number of atoms in the trap is known, say $N$, then $|\phi\rangle = |N\rangle$ (the super atom state) and we have 
\begin{equation}
\mu = \frac{1}{2}\left(\frac{N-1}{ N+1}\right).
\end{equation}
Again, the larger the super atom, i.e., the more particles are present in the trap, the more entanglement we get. As a matter of fact, it seems that almost for any non-trivial superposition of number states $|n\rangle$ (states with zero or one particle only are not entangled) in the trap one can extract some entanglement. This also implies that we can apply our technique to fermionic rather than bosonic systems.

We would like to briefly discuss the validity of the first order perturbation that we used to investigate the temporal behaviour of the particles and the probes. For the perturbative expansion to converge we can assume that $\int_0^Tdtg(t)$ is of the order of $|\alpha|^{-4}$ ($N^{-2}$ for number states), which guarantees that the first order term is $|\alpha|^2$ ($N$) times larger than the second order term. The consequence of this assumption is that the probability to obtain an entangled state of the probes (in the first order perturbation) is proportional to $|\alpha|^{-2}$ ($N^{-1}$). The situation is similar to that of generation of entangled pairs of photons in a parametric-down-conversion process, where most of the time one does not obtain any entangled pair whatsoever.

It may also interesting to consider incoherent mixtures of number states and check how much entanglement one can extract from them. Two very relevant cases are thermal mixture of $|n\rangle$ and a mixture of coherent states describing a situation where the phase of the condensate is not known \cite{coherent-BEC}. In case of the thermal mixture it is tempting to check what is the relation between the critical temperature for the condensation and the amount of entanglement between the probes. Perhaps, the disappearance of entanglement may serve as a signature for the onset of BEC. This would be especially interesting in the case of one dimensional condensates where the critical temperature is not sharply defined via standard methods \cite{ketterle}. This will be investigated in the subsequent papers. 

To summarize, we use two independent probes that locally interact with two spatially disjoint parts of a system consisting of non-interacting boson. Since, at the end of the process the probes become entangled, and all we have used are local interactions between the system and the probes, this forces us to conclude that the bosonic system was spatially entangled to start with. This is a particularly surprising conclusion if the initial state of a system is a coherent BEC with a large number of atoms. The entanglement that we extract is the same as the standard EPR entanglement and can be used for information processing. This adds further support to our previous studies of entanglement in similar systems \cite{ourstuff}. 

We would finally like to discuss briefly how our scheme can be implemented in a realistic scenario. Imagine two spatially extended quantum dots immersed in a BEC made up of phonons created by background molecular vibrations. Then, at sufficiently low temperatures the phonons will largely exist in the lowest vibrational modes which is an effective condensate. The coupling of the dots to the vibrations will be similar to our interaction hamiltonian between the system (phonon BEC) and the probes (dots) \cite{dots}. We therefore expect to observe entanglement between internal excitational degrees of freedom of the two dots for some range of coupling parameters and interaction times.    
It may well be that nature already uses a phonon-to-electron entanglement transfer scheme similar to this to achieve some sort of coherent macroscopic behavior.

{\acknowledgements} DK would like to thank Berge Englert, Christian Miniatura and Thomas Durt for stimulating discussions on entanglement in continuous systems. VV acknowledges financial support from EPSRC, QIPIRC and the EU.

\end{document}